\documentclass[aps, pra, floatfix]{revtex4}
\usepackage{graphicx}
\usepackage{amsmath}
\usepackage{setspace}
\usepackage{color}

\begin{document}
\title{Pair-correlated product speed and angular distributions \\ for the OH+CH$_4$/CD$_4$ reactions: 
Further remarks on their \\ classical trajectory calculations in a quantum spirit}

\author{L. Bonnet,$^{1,2}$
J. Espinosa-Garcia$^{3}$ and J. C. Corchado$^{3}$}

\affiliation{
$^{1}$CNRS, Institut des Sciences Mol\'eculaires, UMR 5255, 33405, Talence, France\\
$^{2}$Univ. Bordeaux, Institut des Sciences Mol\'eculaires, UMR 5255, 33405, Talence, France\\
$^{3}$Departamento de Qu\'imica F\'isica, Universidad de Extremadura, Avenida de Elvas S/N, 06071 Badajoz, Spain} 
\date{\today}
\begin{abstract}
\noindent 
\textcolor{black}{Ten years ago, Liu and co-workers measured pair-correlated product speed and angular distributions for the 
OH+CH$_4$/CD$_4$ reactions at the collision energy of $\sim$ 10 kcal/mol [B. Zhang, W. Shiu, J. J. Lin and K. Liu, 
J. Chem. Phys 122, 131102 (2005); B. Zhang, W. Shiu and K. Liu, J. Phys. Chem. A 2005, 109, 8989].
Recently, two of us could semi-quantitatively reproduce these measurements by performing full-dimensional classical trajectory 
calculations in a quantum spirit on an ab-initio potential energy surface of their own [J. Espinosa-Garcia and J. C. Corchado, 
Theor Chem Acc, 2015, 134, 6 ; J. Phys. Chem. B, Article ASAP, DOI: 10.1021/acs.jpcb.5b04290]. 
The goal of the present work is to show that these calculations can be significantly improved by adding a 
few more constraints to better comply with the experimental conditions. 
Overall, the level of agreement between theory and experiment is remarkable considering the large 
dimensionality of the processes under scrutiny.}  
\end{abstract}
\maketitle
\section{Introduction}
\label{sec:introduction}

The fast-developing velocity map imaging (VMI) technics is increasingly used in molecular beam experiments on gas-phase 
bimolecular reactions \cite{Chandler,Eppink,Lin,Suits,Chichinin,Lee}. These technics make possible the accurate measurement 
of the angle-velocity distribution for a given quantum state of one of the two products. 
This \emph{pair-correlated angle-velocity distribution}  
allows to probe the dynamics of polyatomic chemical reactions at a level of details only possible for 
triatomic reactions before the invention of VMI.

Ten years ago, Zhang \emph{et al.} studied the reactions OH+CH$_4 \longrightarrow$ H$_2$O + CH$_3$ and 
OH+CD$_4 \longrightarrow$ HOD + CD$_3$ (as well as some isotopic variants) by the previous technics 
\cite{Zhang1,Zhang2,Zhang3}. These seven-atom reactions leading to two polyatomic products are among the largest 
bimolecular processes considered to date in this type of experiments. By varying both the collision energy and the 
probed quantum state of the CH$_3$/CD$_3$ radical, they could obtain accurate informations on the subtle way the 
vibrational state distribution of the co-fragment H$_2$O/HOD and the angular distribution may evolve.  

If a qualitative interpretation of such results is possible from simple physical and chemical arguments, 
their quantitative analysis requires the development of theoretical models capable of accurately reproducing them. 
These models can then be used for a detailed understanding of the dynamics.
They can also represent (inexpensive) alternative approaches to predict data needed by atmospherical chemists and 
astrophysicists in order to model planetary atmospheres \cite{Planet} and interstellar clouds \cite{Astro}, respectively.

The title reactions take place in the electronic ground state. The first step of their accurate theoretical description
requires performing ab-initio electronic structure calculations and using fitting procedures to construct 
the corresponding potential energy surface (PES) \cite{PES1,PES2}. This work has recently been performed by two of us
\cite{Joaquin1}. The surface obtained was called PES-2014. Note that the latter takes into account the full 
dimensionality of the system, i.e., 15 (reducing the dimensionality generally makes far easier the construction of 
PESs, but this may induce artificial dynamical behaviours ; treating a molecular collision in its full-dimensionality 
is often preferable). The second step of the simulation consists in moving 
nuclei in the previous PES. Ideally, one would like to do it quantum mechanically, and exactly \cite{Gunnar}. 
For the title reactions, however, the basis sizes necessary to converge quantum scattering calculations are huge, 
thus leading to exceedingly large computation times. 
The alternative is to classically move nuclei, a method traditionally called the quasi-classical trajectory 
method (QCTM) \cite{PR,ST}. However, it is known that in the case where only a 
few vibrational states are available to the products, as is often the case when VMI is used, 
velocity distributions should preferably 
be estimated only from those classical trajectories ending in the products with integer vibrational actions, i.e., 
satisfying Bohr quantization principle in the separated products \cite{LB1}. 
During the last decade, QCTM in this old quantum spirit proved 
to lead to very satisfying predictions as compared to exact quantum scattering calculations or experimental measurements 
\cite{LB1}. This approach has been recently used by two of us in order to reproduce 
the measurements of Liu and co-workers at the collision energy of $\sim$ 10 kcal/mol \cite{Joaquin1,Joaquin2}. 

These QCTM results semi-quantitatively reproduced the experimental evidence, which is already quite satisfying given the size of 
the processes under scrutiny. In particular, the vibrational state distributions of the co-fragment H$_2$O/HOD 
were found to be in good agreement with the experimental ones \cite{Joaquin2}. On the other hand, the
speed distributions associated with a given vibrational state of the co-fragment tend to be broader 
than their measured counterparts \cite{Joaquin2}. Moreover, the pair-correlated angular distributions involve 
significant forward contributions, in contradiction with the experiment \cite{Joaquin1}. The goal of the present note is to 
show that one may reduce these discrepancies by adding a few more constraints in the analysis of QCTM results so as to better 
comply with the restrictions imposed by VMI.
In Section~\ref{II}, these changes 
are detailed within 
the framework of the OH+CD$_4$ reaction, and comparison between its predictions and 
measurements is made for both title processes. Section~\ref{III} concludes.

\section{Method and results}
\label{II}

As we first focus on OH+CD$_4 \longrightarrow$ HOD+CD$_3$, the product quantities involved in the next developments 
are the vibrational energy $E^{vib}_{CD_3}$ of CD$_3$, its rotational energy $E^{rot}_{CD_3}$, the vibrational energy 
$E^{vib}_{HOD}$ of HOD, its rotational energy $E^{rot}_{HOD}$, 
the vibrational actions $a_1$, $a_2$ and $a_3$ associated with the symmetric 
streching, antisymmetric streching and bending vibrational modes of HOD, respectively, the relative translational energy 
$E_{trans}$ between HOD and CD$_3$, the speeds (with algebraic sign) $v$ and $v'$ of CD$_3$ and HOD, respectively, within the center-of-mass system, 
and the scattering angle $\theta$. The total linear momentum within the center-of-mass system is zero. Formally, this is written as  
\\
\begin{equation}
m_{CD_3}v+m_{HOD}v' = 0,
\label{1a}
\end{equation}
\\
where $m_{CD_3}$ is the total mass of CD$_3$ and $m_{HOD}$ is the analogous mass for HOD. Moreover,
\\
\begin{equation}
E_{trans} = \frac{1}{2}m_{CD_3}v^2+\frac{1}{2}m_{HOD}v'^2.
\label{1b}
\end{equation}
\\
From the two previous equations, one gets
\\
\begin{equation}
v=\left[\frac{2 E_{trans}}{m_{CD_3}\left(1+\frac{m_{CD_3}}{m_{HOD}}\right)}\right]^{1/2}.
\label{1}
\end{equation}
\\
 
The vibrational state of HOD is denoted ($n_1$,$n_2$,$n_3$). 
These quantum numbers are associated with the same vibrational modes as the previous actions ($a_1$, $a_2$, $a_3$). 
The corresponding frequencies are $\omega_1$, $\omega_2$ and $\omega_3$ (their values are given in ref.~\cite{Joaquin2}).
The population of state ($n_1$,$n_2$,$n_3$) is denoted $p_{n_1n_2n_3}$ and the contribution to the pair-correlated 
speed distribution $P(v)$ due to state ($n_1$,$n_2$,$n_3$) is denoted $P_{n_1n_2n_3}(v)$. The latter will be normalized to unity.  

In the experiments of Zhang \emph{et al.} \cite{Zhang1,Zhang2,Zhang3}, frozen reagents meet with a collision energy of 
10 kcal/mol (among others) and $P(v)$ is measured for CD$_3$ in its vibrational ground state and its lowest rotational levels 
($N \le 5$) \cite{Lin}. The rotational energy of the probed CD$_3$ is thus lower than $\sim$ 1 kcal/mol. 
Three vibrational states are available to HOD, (1,0,0), (1,0,1) and (2,0,0), by increasing order of energy. 
Therefore,
\\
\begin{equation}
P(v)=p_{100}P_{100}(v)+p_{101}P_{101}(v)+p_{200}P_{200}(v).
\label{O}
\end{equation}
\\

As noted in the introduction, we wish to improve here the agreement between theory and experiment previously
obtained in refs.~\cite{Joaquin1,Joaquin2}. To this aim, we use the same QCTM information as in the previous works 
where full computational details on the calculations can be found. In brief, 10$^5$ trajectories were run on the 
PES-2014, among which 5016 turned out to be reactive. Despite this relatively low number of reactive paths, 
our approach will prove to be efficient, in particular as regards $P(v)$. This is an important asset since 
getting more than a few thousands of reactive trajectories may be computationally expensive for larger direct reactions 
or complex-forming reactions (not to mention processes simulated by on-the-fly calculations). Hence, we did not 
consider performing additional QCT calculations to increase the number of reactive paths. 

Before entering into the details of our approach, a comment on the experimental measurement of $P(v)$ is in order. 
In a hypothetical experiment where both reagents would be in the rovibrational ground state and the collision energy 
would be perfectly controlled, the total energy available to the products would take the same value for the whole set 
of reactions detected. Since $E_{trans}$ is equal to the previous energy minus the quantized internal energy of 
the products, $E_{trans}$ would also be quantized. Consequently, $P(v)$ would be given by a set of Dirac 
peaks with given weights. 
In real experiments, however, one generally observes a continuous density, due to the fact that the previous conditions 
are not fulfilled. For instance, the collision energy and the reagent rotational states are not fully controlled, and 
VMI also introduces some blurring of the peaks. We shall take this into account in the calculation of $P(v)$ in 
Section~\ref{II.B}.

\subsection{Vibrational state populations}
\label{II.A}

\begin{figure}
 \includegraphics[width=1.0\textwidth]{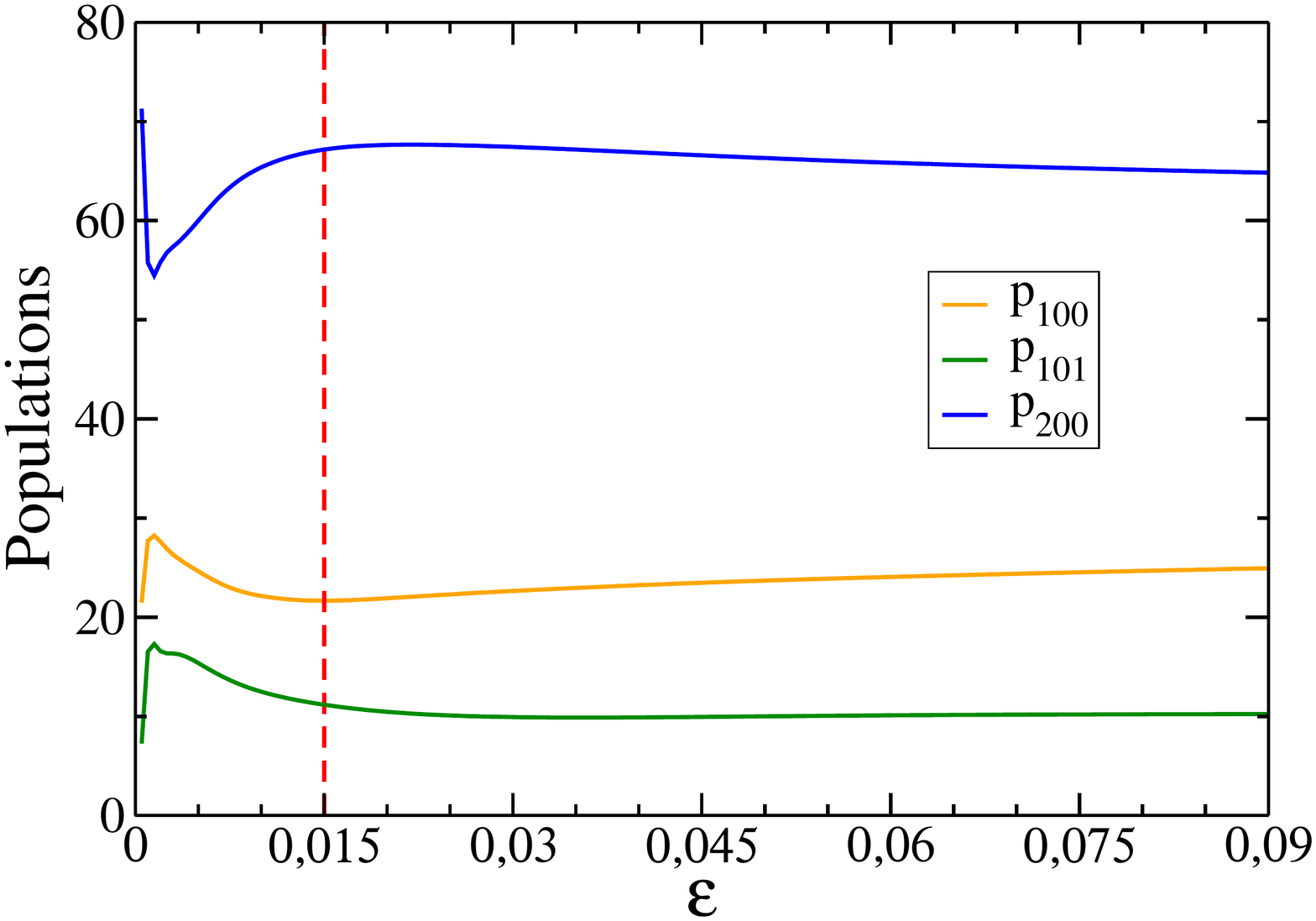}
 \caption{1GB vibrational state populations of HOD (in percentage) in terms of $\epsilon$. \label{figure0}}
\end{figure}

In a first step, we estimate the vibrational state populations of HOD by using nearly the same method as 
in ref.~\cite{Joaquin2}. The zero point energy (ZPE) of CD$_3$ is $E^{0}_{CD_3}=13.8$ kcal/mol. The energy 
of its first excited state is $E^{1}_{CD_3}=14.9$ kcal/mol \cite{Joaquin2},
corresponding to one quantum in the lowest umbrella mode of CD$_3$, i.e., 396 cm$^{-1}$ or 1.1 kcal/mol.
Calling $\delta$ the previous quantum, the intermediate energy between 
the two previous energies is $E^{0}_{CD_3}+\delta/2 =$ 14.35 kcal/mol.
We now assume, in the spirit of standard binning (SB) \cite{LB1}, 
that all the trajectories leading to $E^{vib}_{CD_3} \ge 14.35$ do not contribute to the vibrational ground state of 
CD$_3$ probed by Zhang \emph{et al.} \cite{Zhang1,Zhang2,Zhang3}, and should thus be discarded. 
Over the 5016 reactive trajectories, only 1773 
are then retained. On the other hand, the remaining constraint ($E^{rot}_{CD_3} \le \;\sim 1$ kcal/mol) is ignored in order 
to avoid degrading too much the statistics (see further below). 
Since only three vibrational states of HOD are available, it is preferable, as previously stated, 
to take into account Bohr quantization principle in the analysis of the final trajectory results \cite{LB1}.
For polyatomic molecules, this is usually done by means of the 1GB procedure \cite{Joaquin2,LB1,Gabor,LB2,Gabor2}. 
The 1GB estimation of $p_{n_1n_2n_3}$ is performed as follows: we consider the set of trajectories for which 
$\bar{a}_1=n_1$, $\bar{a}_2=n_2$ and $\bar{a}_3=n_3$, where $\bar{a}_i$ is the nearest integer of $a_i$, $i=$1, 2 or 3. 
In other words, trajectories for which ($a_1,a_2,a_3$) does not belong to the unit cube centred at ($n_1$,$n_2$,$n_3$) 
are rejected. This selection is common to the 1GB and SB procedures. However, contrary to SB which assigns the same 
weight to each trajectory, the 1GB weight is  
\\
\begin{equation}
p_{1GB}(a_1,a_2,a_3) \propto 
exp\left(
-\left[\frac{\sum_{i=1}^3 \omega_i (a_i-n_i)}{\epsilon \sum_{i=1}^3 \omega_i}\right]^2\right)
\label{2}
\end{equation}
\\
within the harmonic treatment of HOD vibrational motions \cite{Joaquin2,LB1,Gabor,LB2,Gabor2}. 
The value to be given to $\epsilon$ is discussed further below. 1GB is an energy-based Gaussian binning.
The sum $\sum_{i=1}^3 \omega_i (a_i-n_i)$ is indeed proportional to the difference between the harmonic classical and 
quantum vibrational energies. When anharmonicities are not negligible, the previous difference can be replaced by the one
between the exact classical and quantum vibrational energies \cite{Gabor2}. This is precisely the case for HOD, and we 
shall thus use for states (1,0,0), (1,0,1) and (2,0,0) the experimental eigenvalues 2723.66, 4100.05 and 5363.59 cm$^{-1}$ 
above the ZPE of HOD, respectively~\cite{Ben}. Note that the set of trajectories used here is still the one for which 
($a_1,a_2,a_3$) belongs to the unit cube centred at ($n_1$,$n_2$,$n_3$). $p_{n_1n_2n_3}$ is obtained by summing the 1GB 
weights over the previous paths and normalizing to unity the set of available populations. In principle, the strict 
application of Bohr quantization requires taking $\epsilon$ as close to 0 as possible \cite{LB1,LB2}. However, making $\epsilon$ 
tend to 0 requires running an infinite number of trajectories, for the percentage of these paths contributing to the populations
is roughly proportional to $\epsilon$. Since a limited number of trajectories are available in practice (1773 here), 
there is a lower bound for $\epsilon$
below which using 1GB makes no sense. In order to illustrate this point, the variation of $p_{100}$, $p_{101}$ and $p_{200}$ 
in terms of $\epsilon$ is displayed in Fig.~\ref{figure0}. 
The vertical red dashed line is defined by $\epsilon = 0.015$.
While $p_{101}$ remains roughly constant when $\epsilon$ decreases from 0.09 down to $\sim$ 0.015, 
$p_{200}$ slowly increases at the expense of $p_{100}$ (the sum of the three populations stays obviously equal to 1). 
In other words, when $\epsilon$ decreases, 
the number of trajectories contributing to $(1,0,1)$ decreases a bit faster than the analogous number for $(2,0,0)$, 
and the same trend is found for $(1,0,0)$ with respect to $(1,0,1)$. Below $\sim$ 0.015, however, a different regime is 
observed, i.e., the rate of decrease becomes increasingly larger for $(2,0,0)$ than for $(1,0,0)$ and $(1,0,1)$ to finally 
get reversed and diverge for $\epsilon$ lower than $\sim$ 0.0015. This sensibility of the populations 
to $\epsilon$ below $\sim$ 0.015 is 
due to the fact that too few trajectories contribute to the statistics. Hence, we assumed that the range of statistical 
confidence lies on the right-side of the red dashed line in Fig.~\ref{figure0} and estimated the 1GB populations at the 
lower limit of this range, i.e., for $\epsilon=0.015$. The resulting populations (rounded to their nearest integer) 
are $p_{200} = 67$, $p_{101} = 11$ and $p_{100} = 22$ $\%$, in satisfying agreement with the experimental 
and previous theoretical populations~\cite{Zhang1,Joaquin2}. 

We wish to emphasize here that the effective number of trajectories 
contributing to the previous populations is 104 for $\epsilon=0.015$ (this number was estimated by counting the trajectories 
for which the right-hand-side (RHS) of Eq.~\eqref{2} is larger than 0.5). Hence, any additional constraint like 
$E^{rot}_{CD_3} \le \;\sim 1$ kcal/mol, or the non violation of CD$_3$ ZPE discussed 
in the next section, would lower this number and make the statistics too poor for the 1GB populations to be realistic.

\subsection{Pair-correlated speed distribution}
\label{II.B}

We now concentrate on the calculation of $P_{100}(v)$, $P_{101}(v)$ and $P_{200}(v)$. 
To deal with the fact that CD$_3$ is selected in its vibrational ground state, we previously limited ourselves to reject
all the trajectories leading to $E^{vib}_{CD_3} \ge E^{0}_{CD_3}+\delta/2$, i.e., 14.35 kcal/mol. If this constraint
allows to discard trajectories contributing to vibrationally excited states of CD$_3$, it leaves unsolved a serious 
issue, i.e., the violation of the ZPE of CD$_3$. Paths leading to $E^{vib}_{CD_3}$ below $E^0_{CD_3}$ 
may indeed artificially overestimate $E_{trans}$ as compared to the experiment. To avoid this, we limit in the following 
the statistics to trajectories for which $E^{vib}_{CD_3} \le E^{0}_{CD_3}-\delta/2$, i.e., 13.25 kcal/mol, 
in addition to applying the previous constraint.
We thus use an energy-based windowing, which is related somehow to the energy-based Gaussian 
binning (1GB) \cite{LB1,Gabor,LB2,Gabor2}. The main difference is that the former procedure also takes into account paths 
for which the nearest integers of the vibrational actions of CD$_3$ are non 0. This is less restrictive, possibly a bit less 
physical, but the determination of the six vibrational actions of CD$_3$ is avoided and we shall see further below that this 
procedure leads to satisfying predictions. 

\begin{figure}
 \includegraphics[width=1.0\textwidth]{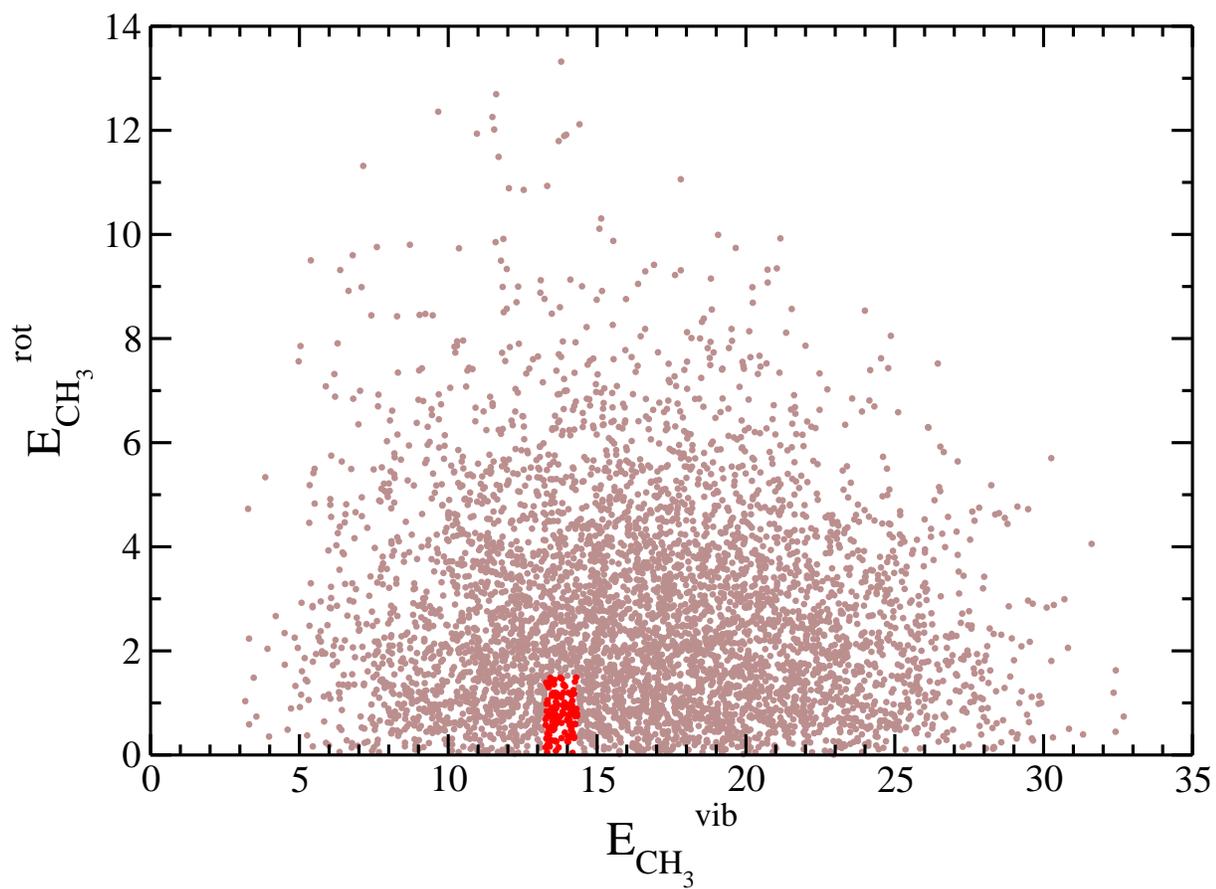}
 \caption{Cloud of points of coordinates ($E^{vib}_{CD_3},E^{rot}_{CD_3}$) for the 5016 reactive trajectories. 
  Grey points correspond to the rejected trajectories while red points correspond to the accepted ones from 
  which the pair-correlated speed and angular distributions will be constructed. \label{figure1}}
\end{figure}

As previously seen, we should in principle reject all the trajectories leading to $E^{rot}_{CD_3}$ larger than $\sim$ 1 kcal/mol, 
but in order to improve the statistics, we slightly shift this limit up to 1.5 kcal/mol. This artificial shift has to be 
as small as possible, since it allows for more energy channeled into the rotational motion of CD$_3$ at the expense of
its translation motion. Consequently, increasing the upper bound of $E^{rot}_{CD_3}$ tends to shift $P_{n_1n_2n_3}(v)$ 
towards small velocities. In Fig.~\ref{figure1}, the cloud of points of coordinates ($E^{vib}_{CD_3},E^{rot}_{CD_3}$) 
is represented for the 5016 reactive trajectories. Grey/red points correspond to rejected/accepted trajectories. 
Only 126 paths turn out to be useful ! Though this number is very small, this will be enough for our purposes.
Note that these trajectories represent only 0.126 $\%$ of the whole set of calculated trajectories, which clearly 
illustrates the fact that simulating polyatomic reactions studied by VMI is numerically very demanding, even 
with QCTM.

\begin{figure}
 \includegraphics[width=1.0\textwidth]{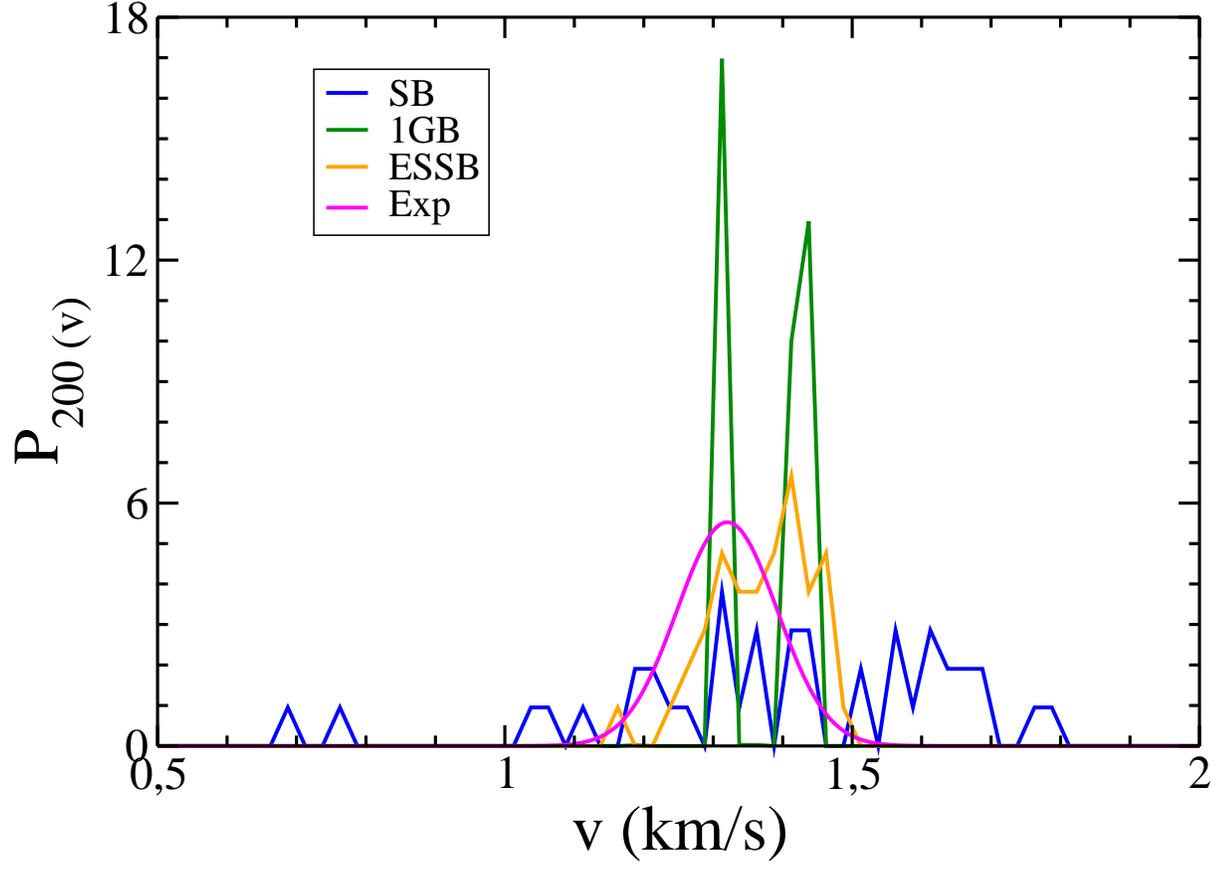}
 \caption{Blue curve: $P_{200}(v)$ obtained from QCTM calculations and the SB procedure.
 Green curve: same as previously with the 1GB procedure. Orange curve: same as previously with the ESSB procedure. 
 Magenta curve: corresponding experimental distribution.  \label{figure2}}
\end{figure}

Like $p_{n_1n_2n_3}$, $P_{n_1n_2n_3}(v)$ should better be calculated in a quantum spirit. In order to illustrate this, 
we first focus on the most probable state (2,0,0) and  
calculate $P_{200}(v)$ in the standard following way ignoring any quantization: we apply the SB procedure, i.e., 
we focus on trajectories for which ($a_1,a_2,a_3$) belongs to the unit cube centred at (2,0,0). 
We then divide in 80 equal intervals the range [0.5, 2.5] comprising the velocities measured in km/s~\cite{Zhang1} 
and just count the number of trajectories per 
interval (the resulting curve is made of points separated by 0.025 km/s, thus leading to a satisfying resolution). 
The resulting SB distribution, normalized to unity, is represented by the blue curve in Fig.~\ref{figure2}, 
to be compared with the magenta experimental curve. Not only does the SB curve involves erratic fluctuations due 
to the small number of trajectories available, but it is also much broader than the magenta curve. We thus calculate 
$P_{200}(v)$ by means of the 1GB procedure, following the same route as above, with the  
difference that trajectories are assigned 1GB weights with $\epsilon = 0.015$ (see Eq.~\eqref{2}) instead of unit weights. 
The resulting distribution is shown in Fig.~\ref{figure2} (green curve). The agreement with the experiment appears to be
very bad. In fact, the effective number of trajectories contributing to $P_{200}(v)$ (number of paths such that the 
RHS of Eq.~\eqref{2} is larger than 0.5) is exceedingly small: only 5 over the 126 available !

To go round this difficulty, we now apply a method already proposed in Sec.~IV of ref.~\cite{LB2}, which amounts to combine 
SB and a translational energy-shifting (ES). We call this alternative procedure ESSB in the following. 
We keep focusing on state (2,0,0).
The set of trajectories used within ESSB is, like within SB, the one for which ($a_1,a_2,a_3$) belongs to the unit cube centred 
at (2,0,0). For simplicity's sake, we first assume that the vibrational modes of HOD are harmonic and treat HOD as a spherical 
top of rotational constant $B$ (deduced from the average of the three moments of inertia of HOD). Hence, the translational energy 
satisfies the identity
\\
\begin{equation}
E_{trans} = E-E^{vib}_{CD_3}-E^{rot}_{CD_3}-\sum_{i=1}^3 \hbar \omega_i \left(a_i+\frac{1}{2}\right)-B J^2
\label{3}
\end{equation}
\\
where $E$ is the total energy available to the products and $J$ is the rotational angular momentum of HOD in $\hbar$ unit.
The 126 trajectories are such that $E^{vib}_{CD_3}$ is close 
to $E^{0}_{CD_3}$, and $E^{rot}_{CD_3}$ to 0 (see the red points in Fig.~\ref{figure1}). We can thus approximate 
$E_{trans}$ by 
\\
\begin{equation}
E_{trans} = E-E^{0}_{CD_3}-\sum_{i=1}^3 \hbar \omega_i \left(a_i+\frac{1}{2}\right)-B J^2.
\label{4}
\end{equation}
\\
When we calculate $P_{200}(v)$ by the SB procedure, we count the number of trajectories for which ($a_1,a_2,a_3$) 
belongs to the unit cube centred at (2,0,0). In nearly 
the same way, if we want to calculate the SB population of HOD in state (2,0,0) and rotational state $j=10$, 
we may count, among the previous set of
trajectories, those for which $J$ belongs to the unit range $[j,j+1]$. Now, the exact quantum value $E^{qm}_{trans}$ 
of the translational 
energy consistent with state (2,0,0) and $j=10$ is equal to the difference between the total energy and  
the energy of the previous state:
\\
\begin{equation}
E^{qm}_{trans} = E-E^{0}_{CD_3}-\sum_{i=1}^3 \hbar \omega_i \left(\bar{a}_i+\frac{1}{2}\right)-B j(j+1).
\label{5}
\end{equation}
\\
This quantity is related to the classical one by
\\
\begin{equation}
E^{qm}_{trans} = E_{trans}+\sum_{i=1}^3 \hbar \omega_i \left(a_i-\bar{a}_i\right)+B[J^2-j(j+1)].
\label{6}
\end{equation}
\\
$J$ is on average equal to $j+1/2$, which is nearly equal to the square root of $j(j+1)$ for not too small 
values of $j$ (these are the most probable ones, due to the degeneracy factor $2j+1$).  
Therefore, $J^2$ is on average equal to $j(j+1)$ and one may neglect the right-most term in the above identity, finally obtaining 
\\
\begin{equation}
E^{qm}_{trans} = E_{trans}+\sum_{i=1}^3 \hbar \omega_i \left(a_i-\bar{a}_i\right).
\label{7}
\end{equation}
\\
The rotational energy not longer appears, so the new translational energy is just 
shifted by the difference between the classical and quantum vibrational energies. 
For a symmetric rotor, the analytical treatment is a bit more
involved, but the final conclusion is the same. For an asymmetric rotor, the problem is far more complex, but Eq.~\eqref{7} 
should still be reasonable. When anharmonicities are not negligible \cite{Gabor2}, which is the case for HOD, 
it is preferable to use the more accurate expression 
\\
\begin{equation}
E^{qm}_{trans} = E_{trans}+E^{vib}_{HOD}-E^{vib\;qm}_{HOD},
\label{8}
\end{equation}
\\
where we recall that $E^{vib}_{HOD}$ is the exact classical vibrational energy of HOD and 
$E^{vib\;qm}_{HOD}$ is its exact quantum mechanical counterpart (see previous section). 
Finally, $P_{200}(v)$ is calculated as previously explained, 
except that $v$ is deduced from $E^{qm}_{trans}$ instead of $E_{trans}$ (see Eq.~\eqref{1}). The resulting distribution, shown 
in Fig.~\ref{figure2} (orange curve), appears to be in much better agreement with the experimental one than the SB and 1GB curves. 
In particular, the former is less irregular and its width quite satisfying. 

\begin{figure}
 \includegraphics[width=1.0\textwidth]{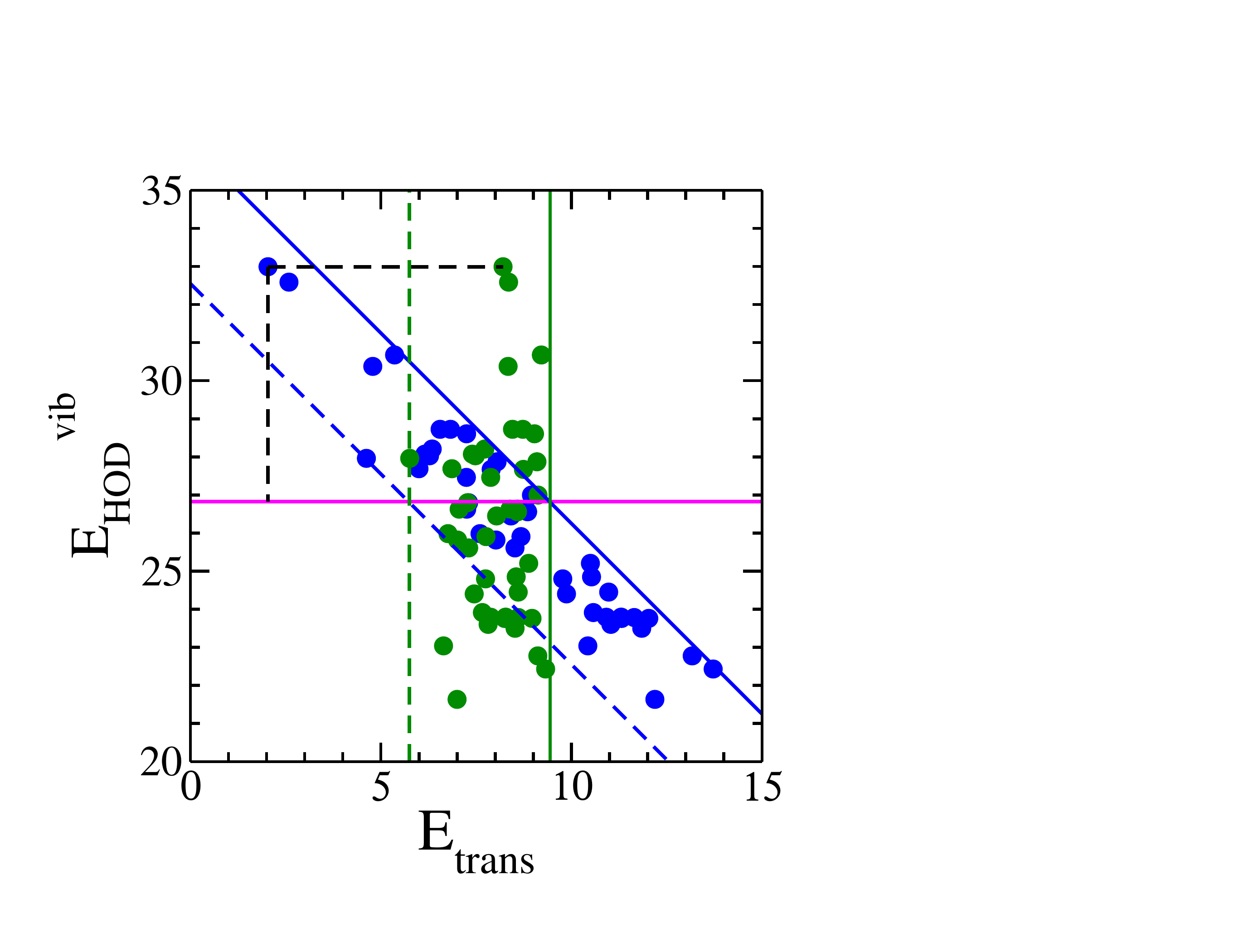}
 \caption{Geometrical transformation associated with the ESSB procedure. See text for explanations.  \label{figure3}}
\end{figure}

In order to explain why the ESSB distribution is narrower than the SB one, we consider the blue circles in 
Fig.~\ref{figure3}, centered at the points of coordinates ($E_{trans},E^{vib}_{HOD}$) obtained from the trajectories 
contributing to state (2,0,0). These points are solution of Eq.~\eqref{4}
(within the assumptions $E^{vib}_{CD_3}=E^{0}_{CD_3}$ and $E^{rot}_{CD_3}=0$), rewritten here as
\\
\begin{equation}
E^{vib}_{HOD}= E-E^{0}_{CD_3}-E_{trans}-E^{rot}_{HOD}
\label{4a}
\end{equation}
\\
($E^{vib}_{HOD}$ has been substituted to the harmonic expression of the HOD vibrational energy while 
$E^{rot}_{HOD}$ has been substituted to $B J^2$). 
The blue solid line in Fig.~\ref{figure3} corresponds to the above identity with $E^{rot}_{HOD}=0$. All the blue circles 
lie below this line since $E^{rot}_{HOD}$ is positive. 
The blue dashed line crosses the blue circle lying at the largest distance from the blue solid line, i.e., 
corresponding to the largest value of $E^{rot}_{HOD}$. This value turns out to be relatively small compared to the maximum 
values available to $E^{vib}_{HOD}$ and $E_{trans}$, thus implying that
the band defined by the two blue lines, where the blue circles lie, is relatively narrow. 
The magenta horizontal 
line is defined by $E^{vib}_{HOD}=E^{vib\;qm}_{HOD}=26.79$ kcal/mol, i.e., the energy of state (2,0,0).
The green circles in Fig.~\ref{figure3} are centered at the points of coordinates 
($E^{qm}_{trans},E^{vib}_{HOD}$). The geometric transformation to go from the blue to the 
green circles is suggested by the two black dashed segments. The horizontal one connects a blue circle 
and a green one related by the previous transformation. According to Eq.~\eqref{8}, the length of the 
horizontal segment, i.e., $(E^{qm}_{trans}-E_{trans})$, is equal to the length of the vertical segment, 
i.e., $(E^{vib}_{HOD}-E^{vib\;qm}_{HOD})$. The transformation applied to the blue solid and dashed lines leads to the 
green solid and dashed lines, respectively. The oblique blue cloud is thus transformed into a vertical green cloud.
On the other hand, the centers of gravity of the two clouds are roughly the same. 
As a consequence, the distribution of $E^{qm}_{trans}$ is compressed with respect to the distribution 
of $E_{trans}$ while the average translational energy is roughly the same for both clouds. This conclusion 
obviously holds for the speed distribution.

The 1GB procedure puts strong emphasis on those blue circles lying in the close neighborhood of the magenta line. 
In Fig.~\ref{figure3}, one clearly sees two groups of circles satisfying such a condition. Note that in this figure,
a blue circle is hidden by a green circle whenever they overlap, which is the case right on the magenta curve where 
the previously discussed transformation reduces to identity. 
This is the reason why only a reduced part of the blue circles can be seen close to the magenta line. 
These points are responsible for the two 
peaks of the green curve in Fig.~\ref{figure2}. Clearly, a continuous distribution of blue points obtained from an infinite 
number of trajectories would lead to a 1GB distribution of $E_{trans}$ with a width equal to the length of the magenta segment
joining the blue lines. Now, this length is also the width of the band defined by the green lines. Consequently, 
while only a small percentage of blue circles contribute to the 1GB distribution (the smaller $\epsilon$, the smaller this 
percentage), 100 percent of the green cicles contribute to the ESSB distribution. Therefore, the strength of the ESSB procedure 
is that it leads to a much larger convergence of the speed distribution in terms of number of trajectories run than the 1GB procedure.  

\begin{figure}
 \includegraphics[width=1.0\textwidth]{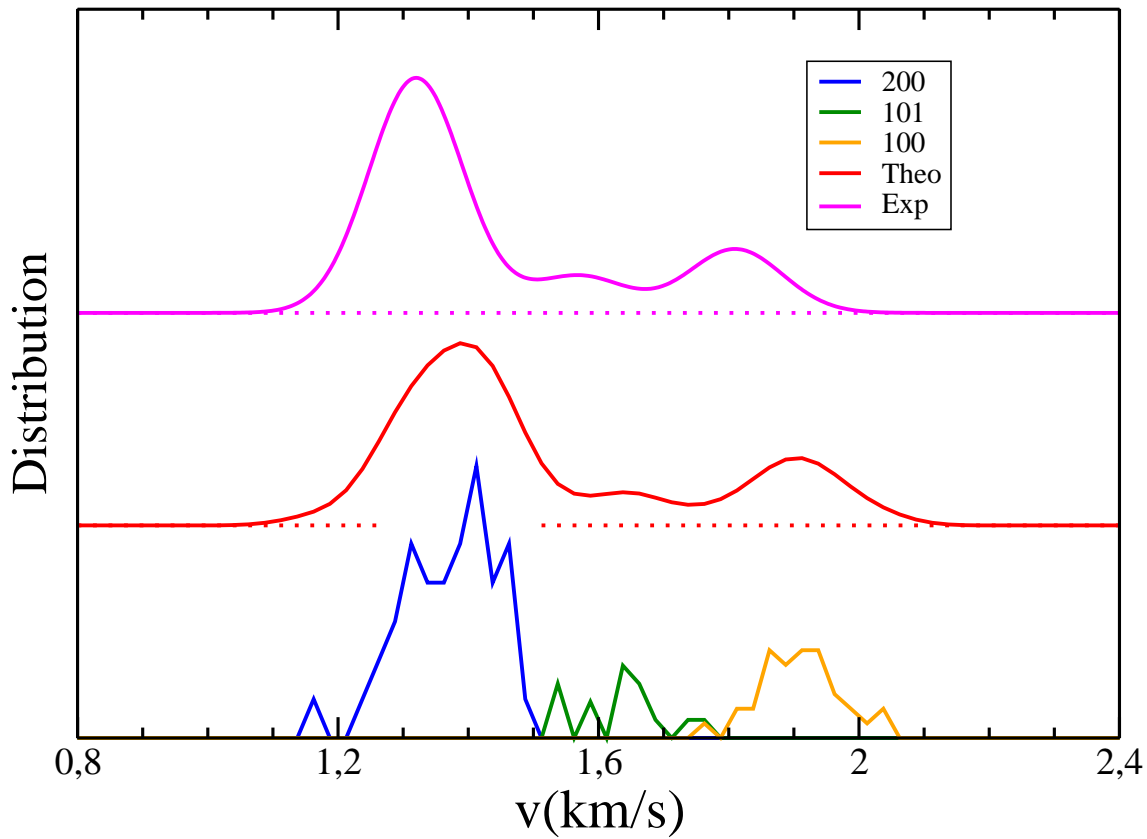}
 \caption{Pair-correlated speed distributions for the OH+CD$_4$ reaction.  
 Blue, green and orange curves: weighted ESSB distributions $p_{200}P_{200}(v)$, $p_{101}P_{101}(v)$
 and $p_{100}P_{100}(v)$, respectively. Red curve: final theoretical result obtained from a 
 convolution of their sum (see Eq.~\eqref{9}). 
 Magenta curve: experimental distribution of Zhang \emph{et al.} (see Fig. 2a in ref.~\cite{Zhang1}). \label{figure5}}
\end{figure}

The remaining distributions $P_{100}(v)$ and $P_{101}(v)$ have been calculated by means of the same ESSB procedure. 
$P_{200}(v)$, $P_{101}(v)$ and $P_{100}(v)$, weighted by the 1GB vibrational state populations, are represented in 
Fig.~\ref{figure5}.
The last step consists in performing the Gaussian convolution 
\\
\begin{equation}
P_c(v)=\int\;dv'\;
\frac{1}{\sqrt{\pi}\eta}exp\left[-\left(\frac{v'-v}{\eta}\right)^2\right]\;
P(v')
\label{9}
\end{equation}
of the total speed distribution
in order to account for its experimental broadening, as previously seen. 
The value of $\eta$ is chosen so as to reproduce as well as possible
the speed distribution at the threshold and cut-off of the experimental distribution. With $\eta$ kept at 0.07,
one obtains the red curve in Fig.~\ref{figure5}, to be compared with the experimental one (magenta 
curve), extracted from Fig.~2a of ref.~\cite{Zhang1}. 

\begin{figure}
 \includegraphics[width=1.0\textwidth]{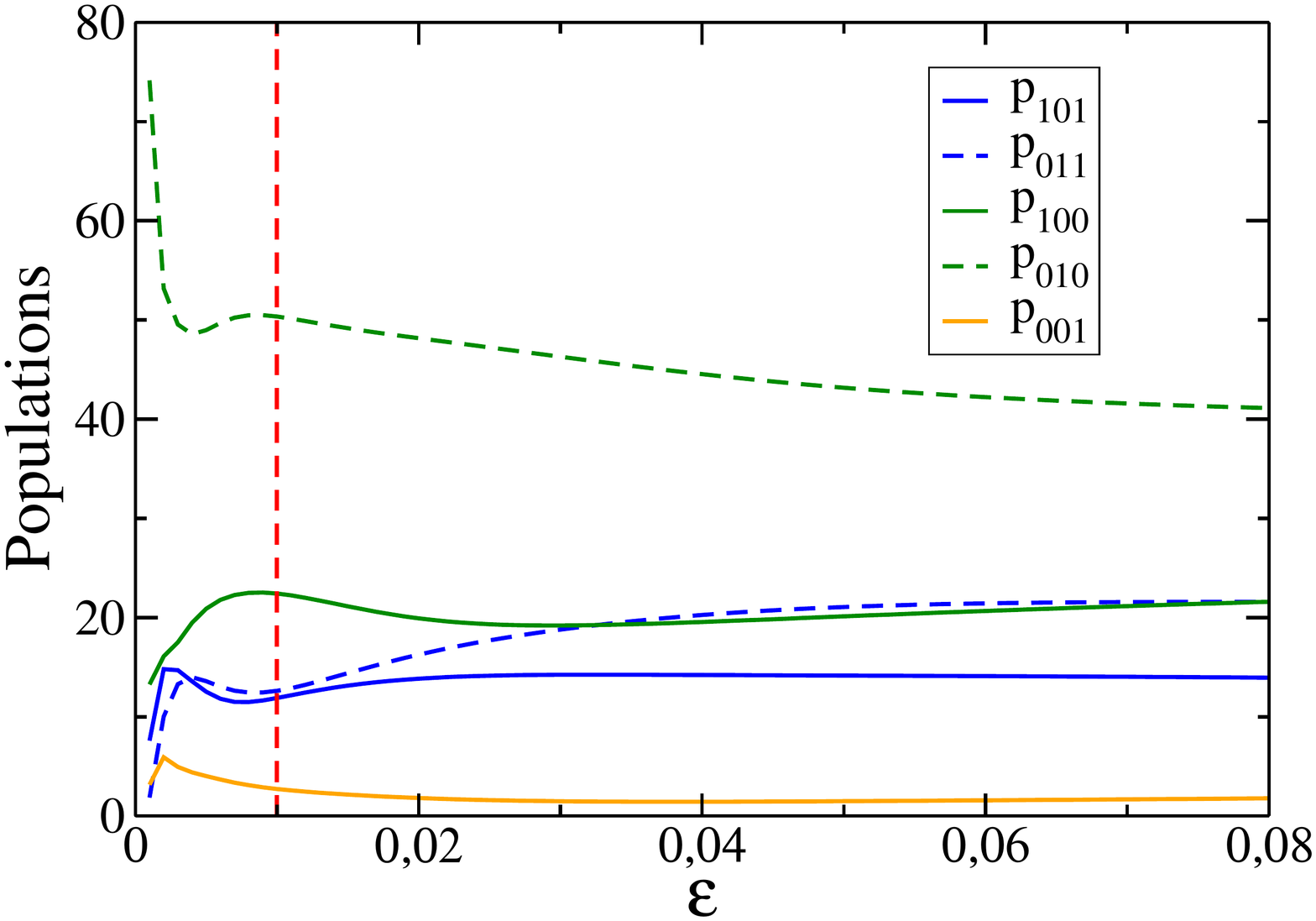}
 \caption{1GB vibrational state populations of H$_2$O (in percentage) in terms of $\epsilon$. \label{figure6}}
\end{figure}

\begin{figure}
 \includegraphics[width=1.0\textwidth]{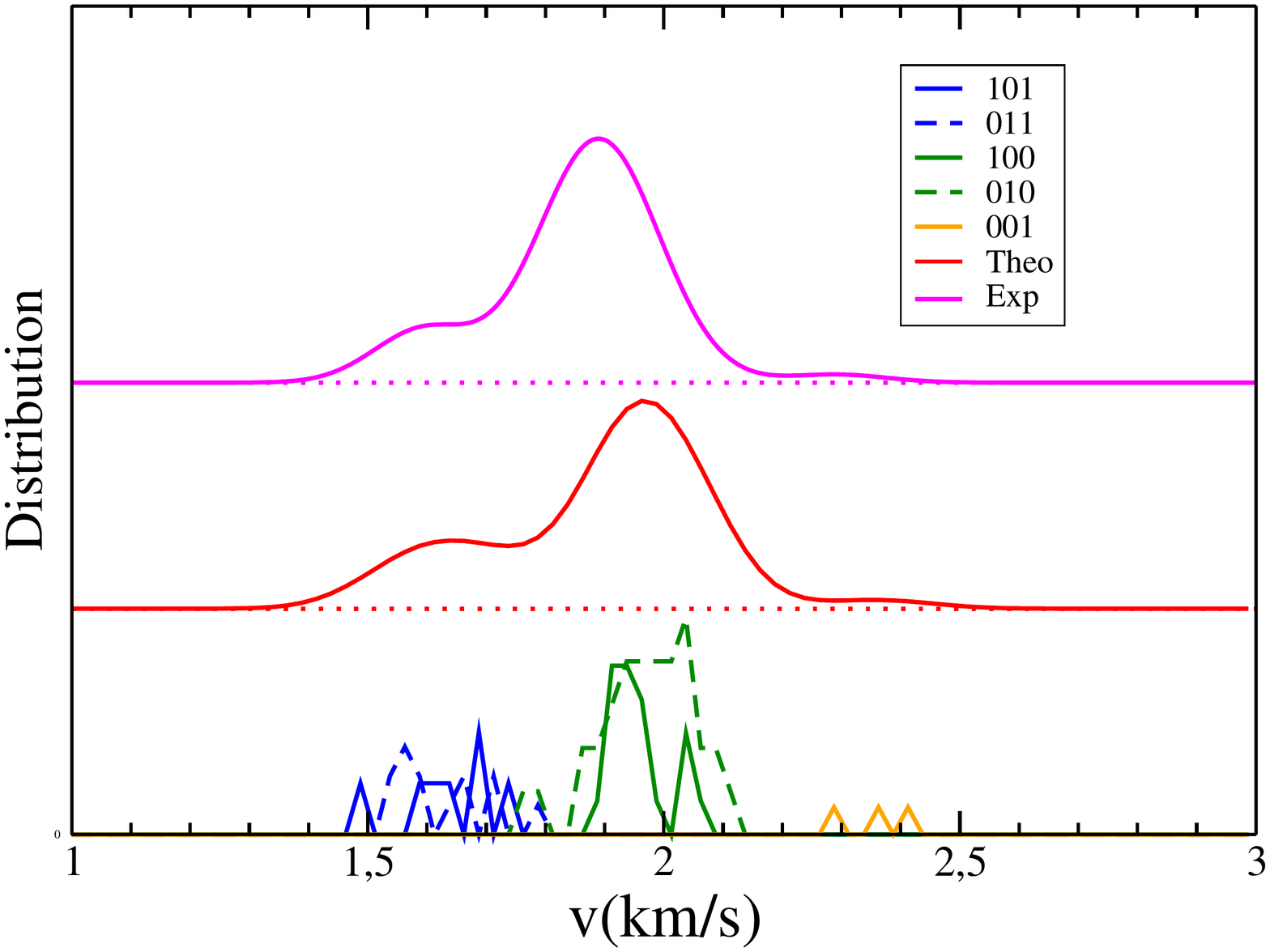}
 \caption{Pair-correlated speed distributions for the reaction OH+CH$_4$. 
 Solid blue, dotted blue, solid green, dotted green and orange curves: weighted ESSB distributions 
 $p_{101}P_{101}(v)$, $p_{011}P_{011}(v)$, $p_{100}P_{100}(v)$, $p_{010}P_{010}(v)$
 and $p_{001}P_{001}(v)$, respectively. Red curve: final theoretical result obtained from a 
 convolution of their sum (see Eq.~\eqref{9}). 
 Magenta curve: experimental distribution of Zhang \emph{et al.} (see Fig. 2a in ref.~\cite{Zhang3}).  \label{figure7}}
\end{figure}

As far as the OH+CH$_4$ reaction is concerned, the following five states of H$_2$O are available: (1,0,1), (0,1,1), (1,0,0), 
(0,1,0) and (0,0,1)~\cite{Zhang3}. We checked that the harmonic normal mode description of water vibrations is very satisfying for 
the energies of the previous states. Consequently, we use Eq.~\eqref{7} for the determination of $E^{qm}_{trans}$.
Apart from that, we apply exactly the same method as for the previous process (with the difference that the ZPE of CH$_3$ 
is 18.8 kcal/mol, and the lowest umbrella frequency is 510 cm$^{-1}$ or 1.5 kcal/mol).
Over 100,000 trajectories run on the PES-2014~\cite{Joaquin1,Joaquin2}, 7525 are found to be reactive, 3535 
contribute to the 1GB populations, and 132 to the pair-correlated speed distribution.
The dependence of the 1GB populations on $\epsilon$ is displayed in Fig.~\ref{figure6}. 
The region of statistical confidence roughly lies on the RHS of the red dashed line, defined by $\epsilon = 0.01$. 
At this frontier, the 1GB populations are equal to 
$p_{101} = 12$, $p_{011} = 13$, $p_{100} = 22$, $p_{010} = 50$ and $p_{001} = 3$ $\%$. 
$P_{101}(v)$, $P_{011}(v)$, $P_{100}(v)$, $P_{010}(v)$ and $P_{001}(v)$, weighted by the previous 
populations, are represented in Fig.~\ref{figure7}. With $\eta$ kept at 0.11, Gaussian convolution of the sum of these densities 
leads to the red curve in Fig.~\ref{figure7}, to be compared with the experimental magenta curve, extracted from Fig.~2a of 
ref.~\cite{Zhang3}. Note that states (1,0,1) and (0,1,1) have similar energies, so the speed distributions associated with these 
states overlap (see the blue curves in Fig.~\ref{figure7}). This comment holds for states (1,0,0) and (0,1,0)
(see the green curves in Fig.~\ref{figure7}). This is the reason why the vibrational populations deduced from the speed 
distributions were previously denoted $p_{01^*1}$ for states (1,0,1) and (0,1,1), $p_{01^*0}$ for states (1,0,0) and (0,1,0) and $p_{001}$
\cite{Joaquin2}. From the present calculations, $p_{01^*1}=25$ and $p_{01^*0}=72$ (see above for the value of $p_{001}$), in 
satisfying agreement with the experimental and previous theoretical populations~\cite{Zhang3,Joaquin2}.

For both processes, the agreement between theory and experiment is remarkable considering the size of the 
reactions under scrutiny as well as the restrictions imposed by the quantization of product vibrations and VMI.
Note that theoretical pair-correlated speed distributions have rarely been published for polyatomic reactions 
(see ref.~\cite{Gabor3} for a recent example in the case of O+CHD$_3$).

\subsection{Angular distributions}
\label{II.C}

In order to obtain $P(\theta)$, we divide the range [0, $\pi$] in 50 equal intervals and count 
the number of trajectories per interval complying with the constraints related to VMI (see previous section).
Then, we use a Legendre polynomial fitting to deduce from the previous numbers the 
pair-correlated angular distributions (20 polynomials are used). 

\begin{figure}
 \includegraphics[width=1.0\textwidth]{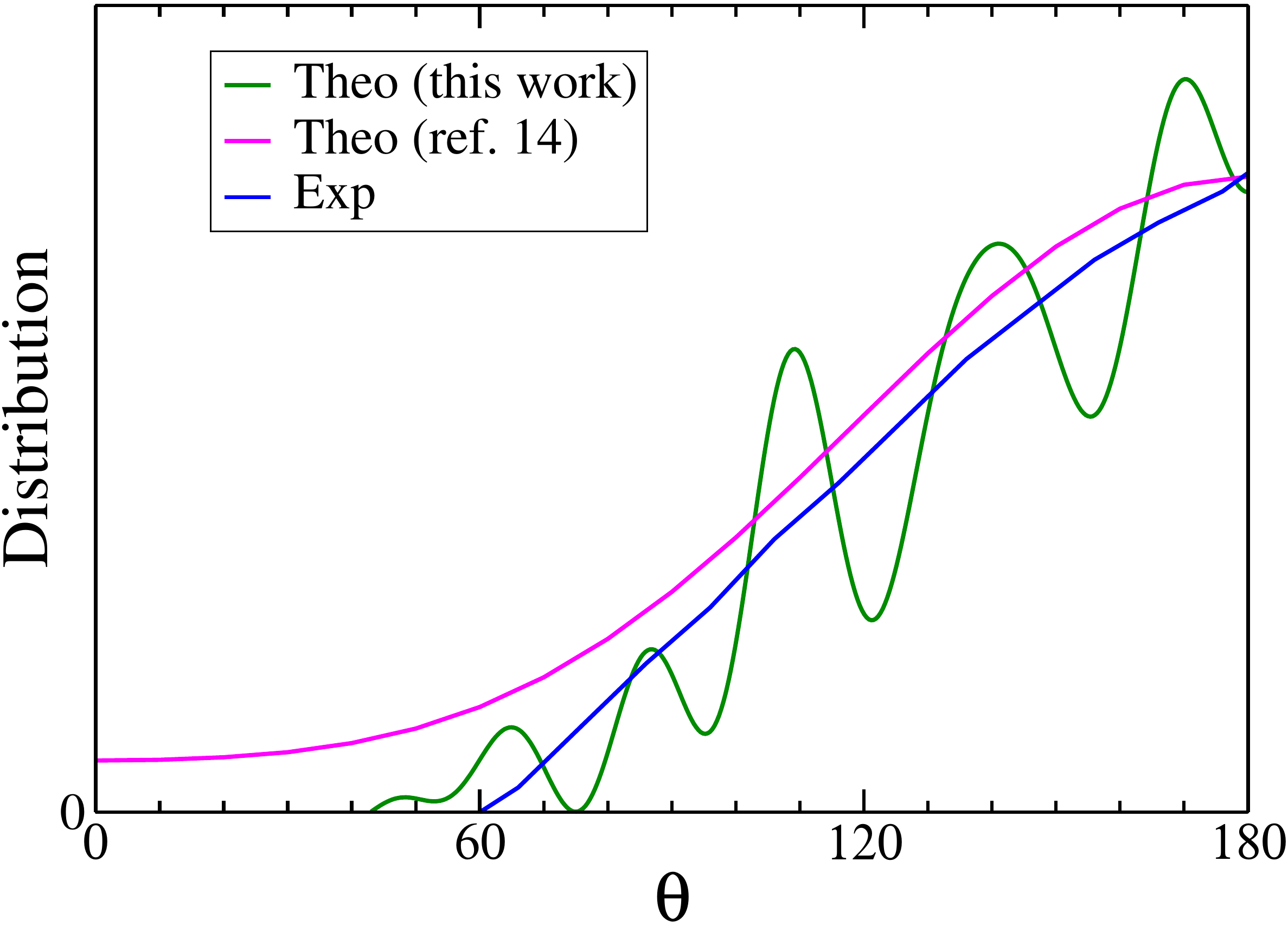}
 \caption{Pair-correlated angular distribution for the reaction OH+CD$_4$. Green curve: present theory. Magenta curve: theory of ref.~\cite{Joaquin1}. 
 Blue curve: experimental results of Zhang \emph{et al.} 
 (see Fig. 3a in ref.~\cite{Zhang1}).  \label{figure8}}
\end{figure}

\begin{figure}
 \includegraphics[width=1.0\textwidth]{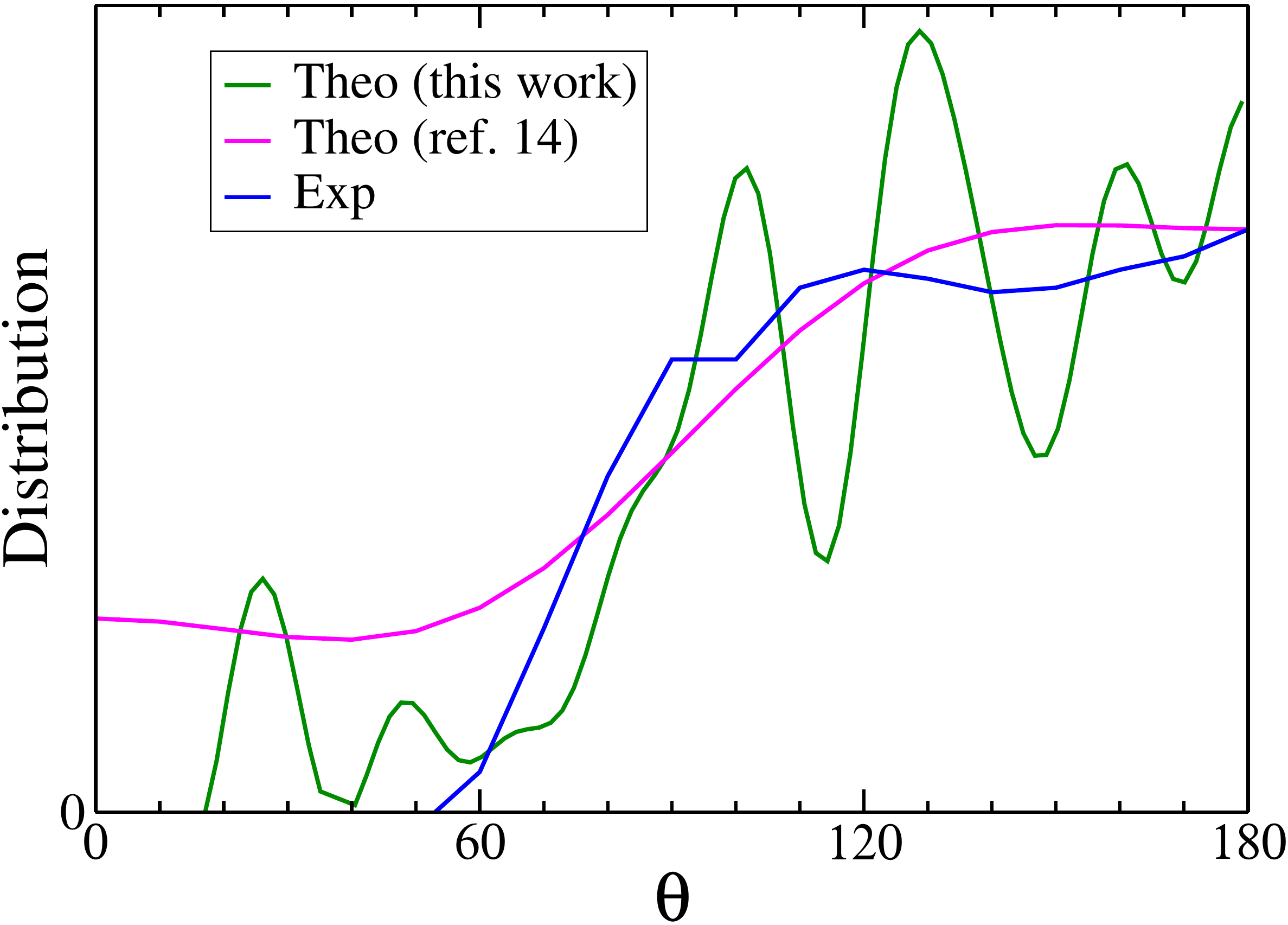}
 \caption{Pair-correlated angular distribution for the reaction OH+CH$_4$. Green curve: present theory. Magenta curve: theory of ref.~\cite{Joaquin1}.
 Blue curve: experimental results of Zhang \emph{et al.} 
 (see Fig. 4a in ref.~\cite{Zhang3}).  \label{figure9}}
\end{figure}

The resulting distributions are represented by green curves in Fig.~\ref{figure8} and Fig.~\ref{figure9} for OH+CD$_4$ 
and OH+CH$_4$, respectively. The magenta curves are extracted from ref.~\cite{Joaquin1} (see Fig. 7 in this work). 
The blues curves are the experimental results, deduced from refs.~\cite{Zhang1} and~\cite{Zhang3}. While 
the magenta and blue curves have been normalized so as to take the same value at $\pi$, 
the green curves have been normalized so as to oscillate around the experimental distribution (for $\theta$ larger 
than $\pi$/2). 

The agreement between the present theoretical results (green curves) 
and the experimental ones (blue curves) appears to be 
satisfying, despite the strong fluctuations observed in the theoretical 
distributions due to the small number of available trajectories (126 and 132 for OH+CD$_4$ and OH+CH$_4$, respectively).
The forward contribution previously reported (see Fig. 7 in ref.~\cite{Joaquin1}) has 
totally disappeared for OH+CD$_4$, and significantly diminished for OH+CH$_4$, thus improving the quality of the predictions.

\section{Conclusion}
\label{III}

The development of the velocity map imaging (VMI) technics~\cite{Chandler} 
over the last twenty years~\cite{Eppink} has made possible the study of polyatomic reactions 
at an unprecedent level of details. For the title processes, for instance, one may get accurate 
data on the way H$_2$O/HOD vibrates and rotates for a given quantum state (or set of closely spaced quantum states) 
of CH$_3$/CD$_3$~\cite{Zhang1,Zhang2,Zhang3}. One may also measure where these products go within the center-of-mass frame. 
If one is able to accurately reproduce these data by 
an adequate theoretical treatment of nuclear motions on a high-quality \emph{ab-initio} potential energy surface, one may then 
use this treatment to get a deep understanding of the dynamics of the processes under scrutiny.  
We have confirmed in this work that the classical trajectory method can be such an approach, provided that Bohr quantization
of product vibration motions is taken into account in the final analysis of the trajectory results~\cite{LB1,Joaquin2,Gabor,LB2,Gabor2}. 
In particular, ways to cope with the drastic reduction of the number of useful trajectories implied by VMI have been proposed.
We plan to apply these ideas to several polyatomic bimolecular reactions and photodissociations for which pair-correlated 
speed and angular distributions have never been theoretically reproduced.

\begin{acknowledgments}
LB is grateful to Prof.~Kopin Liu for clarifying explanations on the cross-beam experiments performed in his laboratory. 
\end{acknowledgments}


\newpage

\newpage

\newpage

\newpage

\newpage

\newpage

\newpage

\newpage

\newpage

\end{document}